\documentclass[twocolumn,aps,pra,superscriptaddress,nofootinbib]{revtex4-1}

\usepackage{graphicx}
\usepackage{dcolumn}
\usepackage{bm}
\usepackage[latin5]{inputenc}
\usepackage{amssymb,amsmath,amsthm,amsbsy,mathptmx}
\DeclareMathAlphabet{\mathcal}{OMS}{cmsy}{m}{n}
\usepackage{tabularx}
\usepackage{xcolor}
\usepackage{mathtools}
\usepackage{oubraces}
\usepackage{dsfont}
\usepackage{graphicx}
\usepackage{rotating}
\usepackage[most]{tcolorbox}
\usepackage{enumerate}
\usepackage{subfloat}
\usepackage{cleveref}
\Crefname{subfigures}{figure}{figures}%
\Crefname{subfigures}{Figure}{Figures}%

\tcbset{colback=green!5!white, colframe=red!75!black,
        highlight math style= {enhanced, 
            colframe=red,colback=red!10!white,boxsep=0pt}
        }

\newcommand*{\coloneqq}{\mathrel{\vcenter{\baselineskip0.5ex \lineskiplimit0pt \hbox{\scriptsize.}\hbox{\scriptsize.}}} =}

\usepackage{hyperref}
\def\bra#1{\mathinner{\langle{#1}|}}
\def\ket#1{\mathinner{|{#1}\rangle}}

\begin{document}

\title{Optimal distillation of quantum coherence with reduced waste of resources}
\author{G\"{o}khan Torun}
\email{torung@itu.edu.tr}
\affiliation{Department of Physics, Istanbul Technical University, Maslak 34469, Istanbul, Turkey}
\affiliation{Centre for the Mathematics and Theoretical Physics of Quantum Non-Equilibrium Systems,
School of Mathematical Sciences, University of Nottingham, Nottingham NG7 2RD, United Kingdom}
\author{Ludovico Lami}
\email{l.lami@nottingham.ac.uk}
\affiliation{Centre for the Mathematics and Theoretical Physics of Quantum Non-Equilibrium Systems,
School of Mathematical Sciences, University of Nottingham, Nottingham NG7 2RD, United Kingdom}
\author{Gerardo Adesso}
\email{gerardo.adesso@nottingham.ac.uk}
\affiliation{Centre for the Mathematics and Theoretical Physics of Quantum Non-Equilibrium Systems,
School of Mathematical Sciences, University of Nottingham, Nottingham NG7 2RD, United Kingdom}
\author{Ali Yildiz}
\email{yildizali2@itu.edu.tr}
\affiliation{Department of Physics, Istanbul Technical University, Maslak 34469, Istanbul, Turkey}

\date{\today}

\begin{abstract}
We present an optimal probabilistic protocol to distill quantum coherence. Inspired by a specific entanglement distillation protocol, our main result yields a strictly incoherent operation that produces one of a family of maximally coherent states of variable dimension from any pure quantum state. We also expand this protocol to the case where it is possible, for some initial states, to avert any waste of resources as far as the output states are concerned, by exploiting an additional transformation into a suitable intermediate state. These results provide practical schemes for efficient quantum resource manipulation.
\end{abstract}

\maketitle


\section*{Introduction}

Over the past three decades quantum entanglement has been identified as one of the main resources that allows us to overcome the intrinsic limits of classical information processing in a distributed setting \cite{Horodecki-review}. It is therefore not surprising that entanglement manipulation is often seen as one of the fundamental tasks in the theory of quantum information. In several cases of practical interest, the goal is that of preparing a target state (e.g., maximally entangled) starting either from many i.i.d.\ copies of the same state \cite{Bennett-distillation, Hayden-EC} or, probabilistically, from a single copy of a known pure state \cite{Popescu-probabilistic, Vidal-probabilistic, Nielsen-LOCC}.
The problem of distilling as much entanglement as possible from a given pure state by means of a probabilistic protocol using local operations and classical communication was considered in Refs. \cite{maxent-set-Plenio,max-distill-Hardy}. Instead of aiming at a single output state, however, one can consider a discrete class of states as targets, namely that formed by all maximally entangled states of any possible local dimension $q$. The protocol given in Refs. \cite{maxent-set-Plenio,max-distill-Hardy} always succeeds in producing one of these states, and a failure occurs only when said local dimension takes the ``trivial'' value $q=1$.

As entanglement of pure states is one of the manifestations of the superposition principle, one can more fundamentally regard the phenomenon of coherent superposition as a valuable resource in its own right. Quantum {\em coherence} plays in fact an essential role in applications to quantum algorithms, quantum metrology, and quantum biology \cite{coherence-review}. To deal with this point of view, a resource theory of quantum coherence has been recently established \cite{Aberg2006, Baumgraz2014, Winter2016, coherence-review, Ringbauer2017}. Coherence distillation is a central task in the resource theory of quantum coherence, and is a subject of very active current investigation \cite{Baumgraz2014, Winter2016, Oneshot1,Oneshot2,OneshotE,OneshotS,Regula2018}.

\begin{figure}[t]
	\centering
	\includegraphics[width=0.44\textwidth]{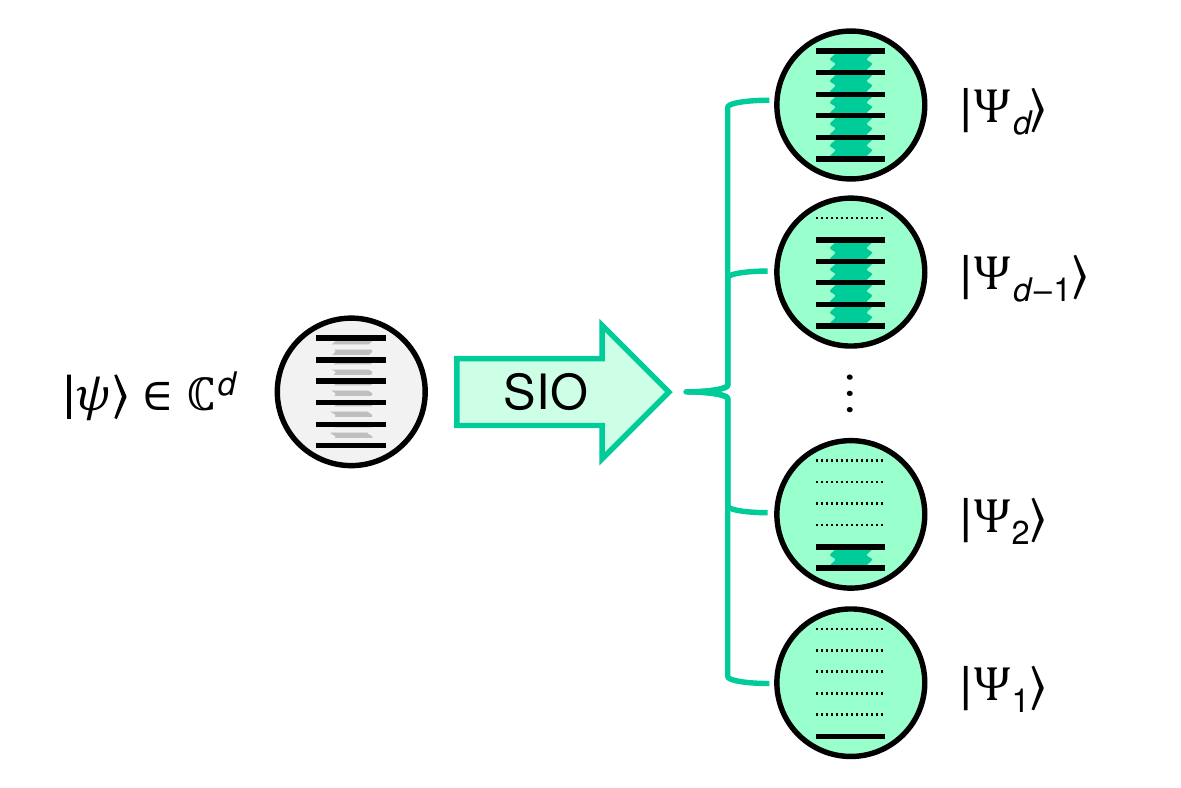}
	\caption{Our strategy solves the coherence distillation problem as follows. We originally have a $d$-level coherent  pure state $\ket{\psi}$. We  perform a strictly incoherent operation (SIO) on the particle  and obtain any of all $q$-level ($q=2,3,\dots,d$) maximally coherent states $\ket{\Psi_q}$, or an incoherent state ($q=1$). The explicit quantum operation used in the protocol is described in the paper.}
	\label{figure1}
\end{figure}

In this paper, we introduce an explicit protocol for coherence distillation via a single strictly incoherent operation
where we originally have a $d$-level coherent input state; see Fig.~\ref{figure1}. This strategy is a counterpart to the entanglement distillation given in Refs. \cite{maxent-set-Plenio,max-distill-Hardy}.
One of the most significant points of this single-step strategy, when we compare it to some common distillation protocols
\cite{multistep-distil1, canonical-distil,Popescu-probabilistic}, is that we can have any of all $q$-level ($q=2,3,\dots,d$) maximally coherent pure states at the end of the measurement process. When compared with the previously available protocols \cite{Popescu-probabilistic}, we see that the failure probability is thus relatively small, and a useful coherent state is almost always produced, unless the incoherent outcome ($q=1$) is obtained. In particular, our protocol is optimal with respect to the distillation of $d$-level maximally coherent states, as the associated probability of success is maximal. We complement our analysis with a quantification of the coherence loss on average in our protocol, and comment on how and for which input states it is possible to modify our strategy, to avoid any waste of resources and always output a state with nonzero coherence.

\section*{Optimal distillation protocol}
To start with, we need to recall the basis-dependent notions of
incoherent and coherent states followed by incoherent operations. Quantum states that are diagonal with respect
to a fixed orthonormal basis $\{\ket{i}\}_{i=1,2,\dots,d}$ are defined as incoherent, and they constitute
a set labeled by $\mathcal{I}$ \cite{Baumgraz2014,coherence-review}. All incoherent states $\rho \in \mathcal{I}$ are of the form
\begin{equation}\label{incoherent-state}
\rho=\sum_{i=1}^{d} p_i \ket{i}\bra{i},
\end{equation}
where $p_i\in [0,1]$ and $\sum_{i}p_i=1$. In addition to this, a finite $d$-dimensional pure coherent state is given by
\begin{eqnarray}\label{psigenform-coh}
\ket{\psi}=\sum_{j=1}^{d} e^{i\theta_j}\psi_j \ket{j}, \quad \big(0\leq\theta_j\leq \pi\big),
\end{eqnarray}
where $\{\psi_j\}_{j=1,2,\dots,d}$ are non-negative real numbers, arranged in nonincreasing order
($\psi_j \geq \psi_{j+1} \geq 0$), and satisfying $\sum_{j=1}^{d}\psi_j^2=1$.
Here, without loss of generality, we can and from now on will assume that $\theta_j=0$ for all $j$, as all these complex phases also can be eliminated by diagonal unitaries, which are always assumed to be free operations in any version of the resource theory of coherence.

We will focus on particular quantum operations for which measurement outcomes are retained as stated in
Ref. \cite{Baumgraz2014}. These quantum operations are defined by Kraus operators $\{K_i\}$ that map incoherent states into incoherent states, i.e., such that
$\sum_{i}K_i^{\dag}K_i=I$ and, for all $i$ and $\rho \in \mathcal{I}$:
\begin{eqnarray}\label{freeoperation}
\rho \rightarrow \rho_i=\frac{K_i\rho K_i^{\dag}}{\text{Tr}[K_i\rho K_i^{\dag}]} \in \mathcal{I}.
\end{eqnarray}
Operations of the form as in Eq. \eqref{freeoperation} in which the Kraus operators satisfy the above condition are known as incoherent operations (IOs) and can be adopted as the
free operations in the context of the resource theory of coherence as defined in Ref. \cite{Baumgraz2014}. A relevant subset of IO is constituted by strictly incoherent operations (SIOs), which are completely positive trace-preserving maps whose Kraus operators $K_i$ satisfy both $K_i\mathcal{I} K_i^{\dagger} \subseteq \mathcal{I}$ and $K_i^{\dagger}\mathcal{I}K_i\subseteq \mathcal{I}$ \cite{Winter2016,SIO-Girolami,SIO-Chitambar,Chitambar2016,Biswas2017,Streltsov2017}.
We will demonstrate that, although SIO have a very limited coherence distillation power when mixed input states are concerned \cite{OneshotS}, they nonetheless suffice for our distillation protocol with pure input states.

We are now ready to analyze the task of one-shot coherence distillation, whose goal is to transform a single copy of the input states given in Eq.~\eqref{psigenform-coh} into a maximally coherent one via (possibly probabilistic) incoherent operations.
The state given in Eq.~\eqref{psigenform-coh} is a $d$-level maximally coherent state for
$\{\psi_i\}_{i=1,\dots,d}\coloneqq\{\frac{1}{\sqrt{d}}, \dots, \frac{1}{\sqrt{d}}\}$:
\begin{eqnarray}\label{max-coherent}
\ket{\Psi_d}=\frac{1}{\sqrt{d}}\sum_{i=1}^{d}\ket{i}.
\end{eqnarray}
An optimal local conversion strategy of bipartite entangled pure states was proposed by Vidal \cite{Vidal-probabilistic}. Adapting those results to the case of coherence distillation, one can obtain the maximal probability of transforming the coherent state $\ket{\psi}$
in Eq.~\eqref{psigenform-coh} to the maximally coherent state $\ket{\Psi_d}$ in Eq.~\eqref{max-coherent} \cite{Du-probabilistic,Liu-max-probability}, which is given by 
\begin{eqnarray}\label{maxprobability}
p(\ket{\psi}\rightarrow \ket{\Psi_d})=\underset{k \in[1,d]}{\min}\left\{\frac{d \sum_{i=k}^{d}\psi_i^2}{d-k+1}\right\}= d \psi_d^2.
\end{eqnarray}

We construct
the explicit Kraus operators to implement the transformations
\begin{eqnarray}
\sum_{i=1}^{d}\psi_i\ket{i} \overset{\text{SIO}}{\longrightarrow} \left\{\Big(p_q, \frac{1}{\sqrt{q}}{\sum_{i=1}^{q}\ket{i}}\Big)\right\}_{q=1,2,\dots,d},
\label{transformations}
\end{eqnarray}
by means of SIOs as defined above. These are given by
\begin{eqnarray}\label{Kraus-d}
K_{q} \coloneqq \sqrt{p_{q}}\left(\frac{1}{\sqrt{q}}\sum_{i=1}^{q}
\frac{\ket{i}\bra{i}}{\psi_i}\right),
\end{eqnarray}
where
\begin{eqnarray} \begin{aligned}
p_d &\coloneqq d\psi_d^2, \\
p_q &\coloneqq q \left( \psi_{q}^2 - \psi_{q+1}^2\right) ,  \quad q=1,2,\ldots, (d-1) .
\end{aligned}
\label{probabilities}
\end{eqnarray}
Note that the operation identified by the above Kraus operators is not only incoherent but also strictly incoherent.
Observe further that the above Kraus operators satisfy the normalization condition $\sum_{i=1}^{d}K_{i}^{\dag}K_{i}=I_d$, implying that they define a legitimate quantum channel. By construction, we have that
\begin{eqnarray}
K_q \ket{\psi} = \sqrt{p_q} \ket{\Psi_{q}} ,\qquad q=1,2,\ldots, d,
\label{measurement-process}
\end{eqnarray}
i.e., such a channel implements the transformations in Eq. \eqref{transformations}. Observe that the success probability of the transformation $\ket{\psi}\to \ket{\Psi_d}$, denoted by $p_d$ and given by Eq. \eqref{probabilities}, achieves its maximal value as given by Eq. \eqref{maxprobability} (see also Ref. \cite{Vidal-probabilistic}). In this sense, the described protocol is optimal.
It is not difficult to verify that the probabilities in Eq.~\eqref{probabilities} correctly satisfy the completeness relation, that is, $\sum_{i=1}^d p_i =1$.
As a result, we initially have the coherent state $\sum_{i=1}^{d}\psi_i\ket{i}$, and after a single-step
measurement process with the given Kraus operators in Eq. \eqref{Kraus-d} we obtain a $q$-level ($q=2,\dots,d$)
maximally coherent state with a certain probability given by Eq.~\eqref{probabilities}. This ensures minimal waste of resources in the distillation protocol, as a useful (albeit of smaller dimension) maximally coherent state is obtained even when the desired outcome is not recorded. Such a feature is explored in more quantitative detail in the following section.



\section*{Coherence loss}
While we know that the degree of coherence can not increase under IOs defined in Eq.~\eqref{freeoperation}, when quantified by suitable coherence monotones \cite{coherence-review}, one may wonder how much coherence is lost on average during our protocol. We adopt the $l_1$ norm of coherence \cite{Baumgraz2014},
a proper quantifier of coherence fulfilling strong monotonicity under IOs, for this study. The $l_1$ norm of coherence of the state
$\ket{\psi}=\sum_{i=1}^{d}\psi_i\ket{i}$ is given by
\begin{eqnarray}\begin{aligned}
C_{l_1}(\rho_{\ket{\psi}}) \coloneqq \Big(\sum_{i=1}^{d}\psi_i\Big)^2-1,
\end{aligned}\end{eqnarray}
and the $l_1$ norm of coherence of the (maximally coherent) state $\ket{\Psi_{q}}=\frac{1}{\sqrt{q}}\sum_{i=1}^{q}\ket{i}$ is given by
\begin{eqnarray}\begin{aligned}\label{l1norm-max-coh}
C_{l_1}(\rho_{\ket{\Psi_{q}}})=q-1, \quad (q=1,2,\dots,d),
\end{aligned}\end{eqnarray}
where $\ket{\Psi_1}=\ket{1}$ is an incoherent state, and, therefore, $C_{l_1}(\rho_{\ket{\Psi_{1}}})=0$.
Combining Eq. \eqref{probabilities} with Eq. \eqref{l1norm-max-coh} we can obtain the average coherence for the output ensemble, given by
\begin{eqnarray}\label{average-coh}
{\bar C}_{l_1}(\rho_{\text{out}})=\sum_{q=1}^{d}p_{q}C_{l_1}(\rho_{\ket{\Psi_{q}}})
=2\sum_{i=1}^{d}(i-1)\psi_i^2.
\end{eqnarray}
Monotonicity (under selective IO on average) yields that
$C_{l_1}(\rho_{\ket{\psi}}) \geq {\bar C}_{l_1}(\rho_{\text{out}})$, i.e.,
$(\sum_{i=1}^{d}\psi_i)^2-1$ $\geq$ $2\sum_{i=1}^{d}(i-1)\psi_i^2$. Thus, the average loss of
coherence for our protocol is found to be
\begin{eqnarray}\label{loss-coh}
C_{l_1}(\rho_{\ket{\psi}})-{\bar C}_{l_1}(\rho_{\text{out}})=\Big(\sum_{i=1}^{d}\psi_i\Big)^2-2\sum_{i=1}^{d}i\psi_i^2+1.
\end{eqnarray}

\begin{figure}[t]
	\centering
	\includegraphics[width=0.4\textwidth]{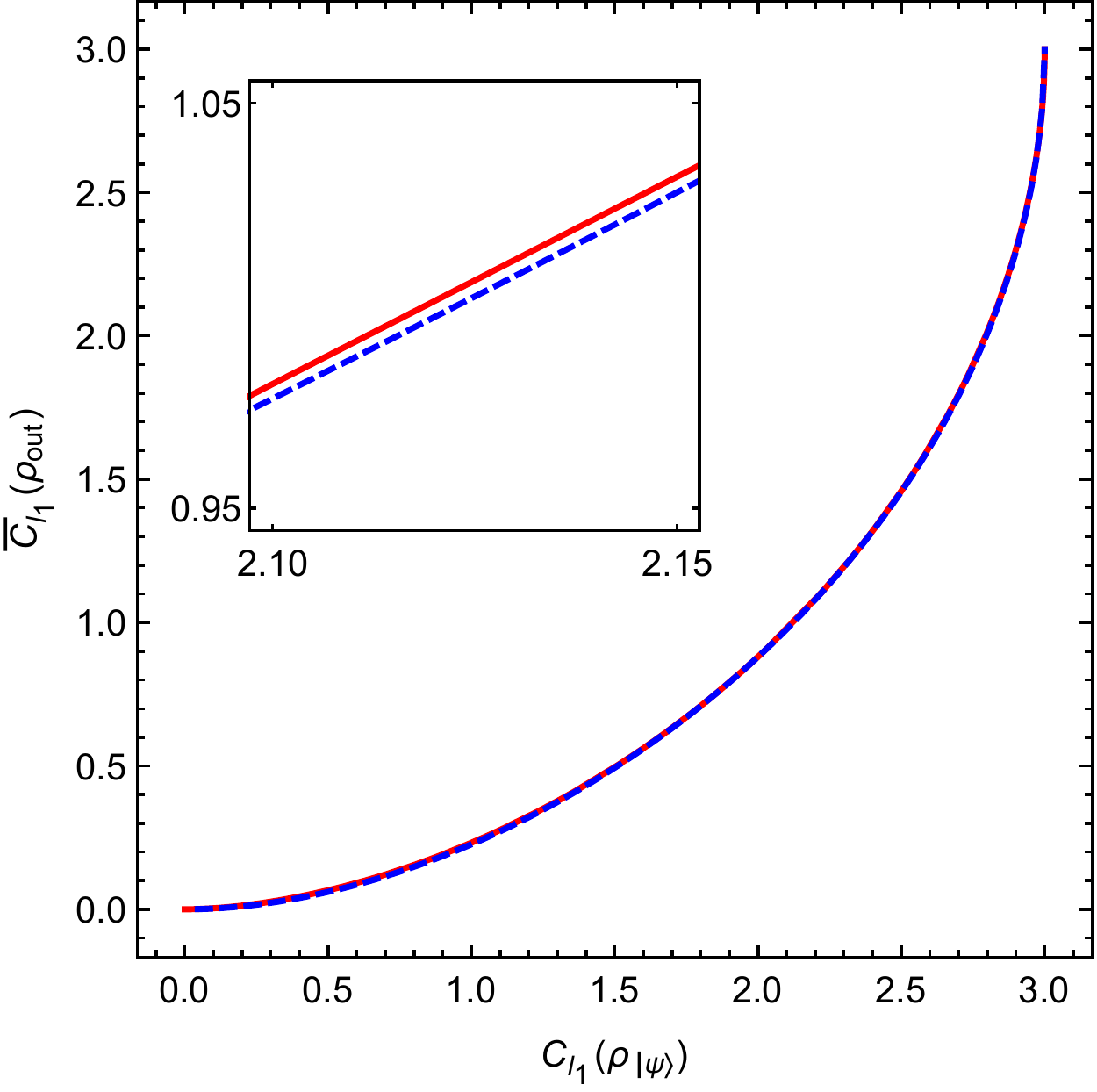}
	\caption{Plot of the output average $l_1$ norm of coherence ${\bar C}_{l_1}(\rho_{\text{out}})$ versus the input coherence $C_{l_1}(\rho_{\ket{\psi}})$ for states in dimension $d=4$. The solid (red) line corresponds to harmonic power states defined by Eq.~(\ref{harmonicst}). The dashed (blue) line corresponds to states maximizing the average coherence loss defined by Eq.~(\ref{loss-coh}), as obtained by solving numerically the corresponding constrained optimization problem. The small difference between the two lines is better seen in the zoomed-in inset. Even if coherence is decreased on average, our protocol always yields a maximally coherent state (in some dimension $q \leq d$) with nonzero probability, apart from the trivial case $q=1$ in which an incoherent state is obtained. All the quantities plotted are dimensionless.}
	\label{figureloss}
\end{figure}

The quantity in Eq.~(\ref{loss-coh}) obviously vanishes when the input is already a $d$-dimensional maximally coherent state, in which case the protocol leaves it invariant with certainty. On the other hand, it can be interesting to investigate classes of states for which there is a large loss of coherence on average during the distillation protocol. One such a class is given by what we may refer to as `harmonic power states', namely, input states $\ket{\psi}$ with coefficients
\begin{eqnarray}\label{harmonicst}
\psi_i=\frac{1}{i^\alpha \ \sqrt{H_d^{(2\alpha)}}}\,,
\end{eqnarray}
where $H_d^{(2\alpha)}$ is the $d^{\text{th}}$ harmonic number of order $2\alpha$, $H_d^{(2\alpha)}=\sum_{j=1}^{d}1/j^{2\alpha}$, with $\alpha \in [0,\infty)$. These states nearly achieve the minimal ${\bar C}_{l_1}(\rho_{\text{out}})$ for a given $C_{l_1}(\rho_{\ket{\psi}})$, as plotted in Fig.~\ref{figureloss} for dimension $d=4$.

\section*{No complete waste of resources}
While the obtained $(d-1)$ outcomes are maximally resourceful states of their corresponding dimension, an incoherent state--waste--is also obtained with a nonzero probability $p_1=\psi_1^2-\psi_2^2$.
The above strategy can be improved so as to avoid complete waste of resources with certainty, provided that the initial state satisfies some mild assumptions on the amount of coherence it contains. Namely, if $\psi_{1}^{2}<1/2$ it is possible to modify the described protocol in such a way as to make $p_{1}=0$, where $p_{1}$ is the probability of outputting an incoherent state (the case $q=1$ in \eqref{measurement-process}). This can be accomplished by first transforming $\ket{\psi}$ into an appropriate intermediate state $\ket{\chi}$, and by finally applying the original protocol to $\ket{\chi}$. The required state $\ket{\chi}$ takes the form
\begin{eqnarray}\label{coh-interm}
\ket{\chi}=\psi_1\sum_{i=1}^{k}\ket{i}+\psi_{k+1}'\ket{k+1}+\sum_{i=k+2}^{d}\psi_i\ket{i},
\end{eqnarray}
where $k>1$ is any integer such that $k\psi_1^2+{\psi'^2_{k+1}}+\sum_{i=k+2}^{d}\psi_i^2=1$ is satisfied for some $\psi'_{k+1}$ subjected to the constraints $\psi_1\geq\psi_{k+1}'\geq\psi_{k+2}\geq\dots\geq\psi_d\geq0$. Using the results in Refs. \cite{Nielsen-LOCC,Condition-Coh-Transfrom-Du,Torung-det}, we know that the transformation $\ket{\psi}\rightarrow\ket{\chi}$ can be performed deterministically. After attaining the temporary state $\ket{\chi}$, we apply the protocol defined by Eq.~\eqref{Kraus-d}, which outputs the ensemble $\ket{\chi} \rightarrow \big\{(p_q, \ket{\Psi_{q}})\big\}_{q=k,\dots,d}$. The entire transformation is then given by
\begin{eqnarray}\label{no-waste-coh}
\ket{\psi} \rightarrow \ket{\chi} \rightarrow \left\{\Big(p_q, \frac{1}{\sqrt{q}}{\sum_{i=1}^{q}\ket{i}}\Big)\right\}_{q=k,\dots,d}.
\end{eqnarray}
It should be highlighted that the probability of obtaining the state $\ket{\Psi_d}$, $p_d=d\psi_d^2$, is still maximum.
Let us discuss a simple example of the above procedure. Consider the initial state $\ket{\psi}=\sqrt{0.35}\ket{1}+\sqrt{0.3}\ket{2}+\sqrt{0.25}\ket{3}+\sqrt{0.1}\ket{4}$ in dimension $d=4$. We can transform this into the temporary state $\ket{\chi}=\sqrt{0.35}\ket{1}+\sqrt{0.35}\ket{2}+\sqrt{0.2}\ket{3}+\sqrt{0.1}\ket{4}$ ($k=2$) with unit probability. Then, the protocol in Eq.~\eqref{Kraus-d} yields the states $\ket{\Psi_4}$, $\ket{\Psi_3}$, and $\ket{\Psi_2}$ with probabilities $0.4$, $0.3$ and $0.3$, respectively.
Ultimately, by the help of a proper intermediate state $\ket{\chi}$ given in Eq. \eqref{coh-interm}, we can obtain an ensemble of maximally coherent $q$-level ($q=k,\dots,d$) states, guaranteeing that all the output states have coherence.

As one can easily notice, this strategy can also be adapted to the entanglement distillation by local operations and classical communication. For the initial bipartite pure entangled state $\ket{\phi}=\sum_{i=1}^{d}\phi_i\ket{ii}$ (Schmidt coefficients are ordered in nonincreasing order as usual), provided that $\phi_{1}^{2}\leq 1/2$ one can find an intermediate state $\ket{\varphi}$ such that
\begin{eqnarray}\label{ent-interm}
\ket{\varphi}=\phi_1\ket{11}+\phi_1\ket{22}+\phi_3'\ket{33}+\sum_{i=4}^{d}\phi_i\ket{ii},
\end{eqnarray}
where $\phi_1\geq\phi_3'\geq\phi_4\geq\dots\geq\phi_d\geq0$. Then, analogously to coherence distillation, using the results in Refs. \cite{Nielsen-LOCC,Torung-det} we can obtain the transformation
$\ket{\phi}\rightarrow\ket{\varphi}$ deterministically in order to avoid producing a separable output with certainty.
The complete transformation is then given by
\begin{eqnarray}\label{no-waste-ent}
\ket{\phi} \rightarrow \ket{\varphi} \rightarrow \Big\{\big(p_q, \ket{\Phi_{q}}\big)\Big\}_{q=2,3,\dots,d},
\end{eqnarray}
where $\ket{\Phi_{q}}=\frac{1}{\sqrt{q}}\sum_{i=1}^{q}\ket{ii}$ and the probability of obtaining the separable state $\ket{\Phi_1}=\ket{11}$ is equal to zero.
Here, while the probability of getting $\ket{\Phi_2}$ and $\ket{\Phi_3}$ increases ($p_{2}=2(\phi^2_1-\phi'^2_3)$) and decreases ($p_{3}=3(\phi'^2_3-\phi_4^2)$), respectively, the other probabilities $p_{m}$ of getting $\ket{\Phi_m}$ ($m=4,5,\dots,d$) remain unchanged.
It is always possible to find an intermediate state $\ket{\varphi}$ of the form \eqref{ent-interm} for the initial states such that $\phi_3^2-\phi_4^2 \geq \phi_1^2-\phi_2^2$.
Therefore, if the initial entangled bipartite states $\sum_{i=1}^{d}\phi_i\ket{ii}$ satisfy this relation, our results ensure that both the transformations given in Eq.~\eqref{no-waste-ent} can be implemented and hence that no waste of entanglement resources is achieved when only the set of the output states are considered.

Another point that needs to be discussed pertains to the largest amount of distilled entanglement. It is given by
\begin{eqnarray}
{\langle E \rangle}_{\text{max}}=\sum_{j=1}^{d}(\lambda_j-\lambda_{j+1})j\ln j,
\end{eqnarray}
for the state $\sum_{j=1}^{d}\sqrt{\lambda_j}\ket{jj}$ \cite{maxent-set-Plenio,max-distill-Hardy}.
Considering the state $\ket{\phi}=\sqrt{0.35}\ket{11}+\sqrt{0.3}\ket{22}+\sqrt{0.25}\ket{33}+\sqrt{0.1}\ket{44}$ as the initial bipartite entangled state, one can transform it into the intermediate state $\ket{\varphi}=\sqrt{0.35}\ket{11}+\sqrt{0.35}\ket{22}+\sqrt{0.2}\ket{33}+\sqrt{0.1}\ket{44}$ deterministically.
Then, the largest amount of distilled entanglement is found to be $1.11821$ and $1.09205$ for the states $\ket{\phi}$ and $\ket{\varphi}$, respectively.
Thus, although the entire transformation $\ket{\phi}\rightarrow\ket{\varphi}\rightarrow\{p_q,\ket{\Phi_{q}}\}_{q=2,\dots,d}$ may provide no complete waste of resources, it may lead to a decreased amount of the largest distilled entanglement.
This is resulting from the increase (decrease) of the probability of obtaining lower (higher) dimensional maximally entangled states.

\section*{Conclusion} In this paper, we presented a simple, practical and efficient strategy for optimal one-shot distillation of quantum coherence from pure input states of arbitrary dimension. The key advantage of our protocol lies in its ability to provide a single map to obtain all $q$-level ($q=2,3,\dots,d$) maximally coherent pure states starting from  a $d$-level coherent input pure state, as illustrated in Fig.~\ref{figure1}. In this way, useful degrees of coherence resource are ``recycled'' even when the maximally resourceful $d$-dimensional state is not obtained.

The probability of success, defined by the outcome $q=d$, is maximal, confirming optimality of the protocol. On the other hand, our protocol only fails when the trivial outcome $q=1$ is obtained, in which case no resource is distilled. This makes our protocol preferable to conventional distillation protocols such as the one in Ref. \cite{Popescu-probabilistic}, which has instead a higher failure probability and produces no useful output in case the desired maximally resourceful output is not obtained. We furthermore showed how to modify the protocol into a two-step strategy which completely nullifies the failure probability, leading to no waste of coherence in the outputs; this is possible for a subclass of input states that we characterize. Our strategy can also be adapted to entanglement distillation.

A further generalization of our scheme and of the seminal works in Refs. \cite{maxent-set-Plenio,max-distill-Hardy} to other quantum resource theories \cite{QRT2018}, beyond coherence and entanglement, would be a worthwhile direction for future investigation.


\begin{acknowledgments}
G.T. acknowledges financial support from the Scientific and Technological Research Council of Turkey (TUBITAK). L.L. and G.A. acknowledge financial support from the European Research Council (ERC) under the Starting Grant GQCOP (Grant No.~637352).
\end{acknowledgments}


\end{document}